\documentclass[12pt]{iopart}

\usepackage{iopams}
\usepackage{bm}

\begin{document}

\title[Short-range plus zero-range  potentials]
{
On  Green's functions for Hamiltonians with potentials possessing singularity at the origin:
application to the zero-range potential formalism
}

\author{S L Yakovlev, V A Gradusov}

\address{ Department of Computational Physics, St Petersburg State University, 198504, St Petersburg, Russia}

\ead{yakovlev@cph10.phys.spbu.ru} \ead{vitaly.gradusov@gmail.com}

\begin{abstract}
We evaluate the short-range asymptotic behavior of Green's function for a Hamiltonian when its potential energy part has
an inverse power singularity at the origin.
The analytically solvable case of sharply screened Coulomb potential is considered firstly. For this potential the additional
logarithmic singular term has been found in the short-range asymptote of the Green function as in the case of the pure Coulomb
potential. The case of a short-range potential
of an arbitrary form with inverse power singularity is treated on the basis of the integral Lippmann-Schwinger equation.
It is shown that, if the singularity is weaker than the Coulomb one, the Green function has only standard singularity.
For the case of $r^{-\rho}$ singularity of the potential with $1\le\rho<2$ the additional singularity in the asymptotic
behavior of the Green function appears.
In the case of $\rho=1$ the additional logarithmic singularity has the same form as in the case of the pure Coulomb potential.
In the case of  $1<\rho<2$
the additional singularity of the Green function has the form of the polar singularity $r^{-\rho+1}$.
These results are applied for extending the zero-range potential formalism on Hamiltonians with singular
potentials.
\end{abstract}


\submitto{\JPA}

\maketitle

\section{Introduction and formulation of the problem}
\label{intro}

In this paper we concentrate on those  attributes of Green's function $G(z)$ of a Hamiltonian $H$, which are needed for evaluating the short-range asymptotic behavior of the Green function in the configuration space. As an operator $G(z)$ is
defined as $G(z)=(H-z)^{-1}$ for $z\in \mathbb{C} $.  The Hamiltonian is assumed to have the
form
\begin{equation}
H=-\Delta + V(\bm{r}),
\label{H}
\end{equation}
where $\Delta$ stands for the Laplacian over $\bm{r}\in \mathbb{R}^3$. While the three-dimensional configuration space is considered,
the case of arbitrary dimension $d>1$ can be treated by similar methods.

 The potential energy term $V(\bm{r})$ is supposed to represent a short-range potential, i.e. it is a real-valued smooth function for all $r \equiv |\bm{r}|>0$ and it vanishes asymptotically as  
$V(\bm{r})\propto r^{-1-\delta}, \ \ \delta>1$
\footnote{This condition for $\delta$ can be weakened up to $\delta>0$. However  we will use the stronger condition $\delta >1$ since it guaranties the absolute convergence of integrals we deal with  in the paper below. } when $ r \to \infty $.
More precisely, we assume that there exists a constant $C>0$ such that the inequality
\begin{equation}
|V(\bm{r})| \le C(1+r)^{-1-\delta}, \delta>1
\label{Vbound}
\end{equation}
holds for all $\bm{r}$  except of a small neighborhood of the origin $\bm{r}=0$.
The specific concern of the paper is the singular short-range behavior of the potential 
$
V(\bm{r})\propto r^{-\rho}
$
 , when $r\to 0$. More precisely, we assume that the potential $V(\bm{r})$ in a neighborhood of the origin $\bm{r}=0$ can be represented as
\begin{equation}
V(\bm{r})=  r^{-\rho} W(\bm{r}),
\label{V at or}
\end{equation}
where $W(\bm{r})$ is a smooth bounded  function with  the finite limit
\begin{equation}
\lim_{r\to 0} W(\bm{r})=V_0.
\label{W at or}
\end{equation}
In what follows this class of potentials will be referred to as the $\mathfrak{V}(\rho,\delta)$ class.
For Hermiticity of $H$ it is sufficient to require $\rho<2$ and therefore it will be assumed throughout the paper that this inequality is fulfilled.

The short-range asymptote of the Green function plays the decisive role in the zero-range potential formalism \cite{albev}.
In our recent paper \cite{yak12} by studying the Green function  we have shown that the zero-range potential has to be modified from the standard form, if it is constructed for the particles interacting by the Coulomb potential $V^{\cal C}(r)=V_0 r^{-1}$. This modification is represented in the short-range behavior of the solution $\phi$ of the Schr\"odinger equation
\begin{equation}
\label{wfasym}
 \phi
\sim \frac{\alpha}{4\pi} \left[ 1/r + V_0\log(r) \right] + \beta, \ \ r\to 0
\end{equation}
by the logarithmic singularity that is additional to the standard  $r^{-1}$  one.
The modification appears in fact as the result of the interplay between singularities of the Coulomb potential and the zero-range potential.
Before treating the general case we will deal with the sharply screened Coulomb potential $V^{\cal C}_R(r)=V^{\cal C}(r)\theta(R-r)$, where $R>0$ is a screening radius and $\theta$-function is defined as $\theta(t)=1\, (0)$  when $t\ge 0\, (<0)$.
We consider this particular case of the screened Coulomb potential
$V^{\cal C}_R$, which is also analytically solvable \cite{yak-screened-coulomb} as $V^{\cal C}$,  in order to emphasize
that only the short-range behavior of the Coulomb potential is responsible for the effect of that interplay and therefore the long-range behavior
of the tail of the  Coulomb potential does not affect the zero-range potential structure.

One of the approaches for constructing the zero-range potential in the general case of $V\in \mathfrak{V}(\rho,\delta)$ consists in inserting a delta-functional term into the Schr\"odinger
equation \cite{yak12}
\begin{equation}
\label{SEVdelta}
\left[-\Delta+V(r)-k^2\right]\phi(\bm{r},\bm{k})+\lambda \delta(\bm{r})\beta=0,
\end{equation}
where $\lambda$ is a coupling constant and $\beta$  is actually a linear functional of $\phi$ \cite{albev}.
Then the solution of (\ref{SEVdelta}) can be given by the Lippmann-Schwinger integral equation
\begin{equation}
\phi(\bm{r},\bm{k}) = \phi_0(\bm{r},\bm{k}) - \lambda \int \mbox{d}\bm{r}'\, G^{+}(\bm{r},\bm{r}',k^2)\delta(\bm{r}')\beta.
\label{LS}
\end{equation}
Here by $G^{+}(k^2)$ we denote $\lim_{\epsilon\to 0}G(k^2+\mbox{i}\epsilon)$ and this notation will be used
systematically throughout the paper. In (\ref{LS}) the function $\phi_0$ is the solution to the equation
\begin{equation}
\label{SEV}
\left[-\Delta+V(r)-k^2\right]\phi_0(\bm{r},\bm{k})=0,
\end{equation}
obeying the asymptotic boundary condition
\begin{equation}
\phi_0(\bm{r},\bm{k})\sim \exp{(\mbox{i} \bm{k}\cdot \bm{r})} + A r^{-1}\exp{(\mbox{i} kr)}
\label{asym bc}
\end{equation}
as $r\to \infty$. The dot-product here and
further means the scalar product of vectors in $\mathbb{R}^3 $.

 The integration in (\ref{LS}) is performed easily
due to the delta-function which yields
\begin{equation}
\phi(\bm{r},\bm{k}) = \phi_0(\bm{r},\bm{k}) - \lambda G^+(\bm{r},0,k^2)\beta.
\label{psi-LS}
\end{equation}
As it will be shown below the limit of $\phi_0$ as $r\to 0$ is finite for $V\in\mathfrak{V}(\rho,\delta) $ and therefore
the non trivial short-range asymptotic of $\phi$ is completely determined by Green's function term in (\ref{psi-LS}).  Hence,  from (\ref{psi-LS}) it is seen that the principal features of the zero-range potential formalism
follow from the short-range behavior in $\bm{r}$  of the Green function $G^+(\bm{r},0,k^2)$
as it takes place in the case of regular potentials \cite{albev}.
In consecutive sections we will evaluate  the respective asymptote of the Green function and as the result the zero-range potential will
be constructed.

The paper is organized as follows. In section 2 we derive the closed form representation for the Green function of the Hamiltonian
 with sharply screened Coulomb potential $V^{\cal C}_R$ and infer the asymptote from it.
As to the best of our knowledge, this representation for $r,r' <R$ has been
obtained  here for the first time. The section 3 is devoted to studying the general case of potentials from $\mathfrak{V}(\rho,\delta)$.
In the section 4  the asymptotes of the Green function are used for evaluating the short-range behavior of the solution
of (\ref{SEVdelta}) and establishing the zero-range potentials for different values of $\rho$. Also in this section  the respective pseudo-potentials are constructed.
The section 5 gives concluding remarks.


\section{Green's function for Screened Coulomb potential}
\label{greenf}

 The Green function is defined by the solution to the inhomogeneous equation
\begin{equation}
\label{defin}
 \left[ -\Delta + V^{\cal C}_R(r) -k^2 \right] G^+_R(\bm{r},\bm{r'},k^2) = \delta(\bm{r}-\bm{r'}).
\end{equation}
One of the convenient ways for constructing the Green function for radial potentials
is the use of the partial wave decomposition~\cite{newton}. The Green function is represented then  as the series in terms of Legendre polynomials
$P_{\ell}$
\begin{equation}
\label{series}
 G^+_R(\bm{r},\bm{r'},k^2) = \frac1{4\pi} \sum_{\ell= 0}^{\infty} (2\ell+1) \frac{G_{R\ell}(r,r',k^2)}{rr'}
 P_\ell(\bm{\hat{r}} \cdot \bm{\hat{r}'}) ,
\end{equation}
where $\bm{\hat{r}}=\bm{r}r^{-1}$. 
The partial Green function $G_{R\ell}$ obviously obeys
the one-dimensional equation
\begin{equation}
\label{e-set}
 \left[ -\frac{\mbox{d}^2}{\mbox{d} r^2} + \frac{\ell(\ell+1)}{r^2} +V^{\cal C}_R(r) -k^2 \right] G_{R\ell}(r,r',k^2) = \delta(r-r')
\end{equation}
with natural boundary condition
$G_{R\ell}=0$ as $r=0$. The radiation boundary condition as $r \to \infty$ requires the outgoing wave asymptote
 $G_{R\ell} \propto \exp(\mbox{i}kr-\mbox{i}\pi \ell/2)$.

The Green function $G_{R\ell}$ can be  constructed following the standard procedure \cite{tit}
\begin{equation}
G_{R\ell}(r,r',k^2)= -\frac{u_{\ell}(r_<)v_{\ell}(r_>)}{W(u_\ell,v_\ell)},
\label{Green's uv}
\end{equation}
where $r_> = \max\{r,r'\}$, $r_< =\min\{r,r'\}$ and $W(u_\ell,v_\ell)$ means the Wronskian of solutions to the equation
\begin{equation}
\label{SE-uv}
 \left[ -\frac{{\mbox d}^2}{{\mbox d}r^2} + \frac{\ell(\ell+1)}{r^2} + V_R(r) -k^2 \right] u(r) = 0.
\end{equation}
The particular solutions $u_{\ell}$ and $v_{\ell}$ should be defined by boundary conditions $u_{\ell}(0)=0$, and $v_{\ell}(r)\to \exp\{\mbox{i}kr-\mbox{i}\pi\ell/2\}$ as $r\to \infty$.
The exact representations for both $u_\ell$ and $v_\ell$ depend on whether the coordinate $r$ is in or out the interval $0 <r \le R$. These representations can be obtained
by the matching technique as in \cite{yak-screened-coulomb}.
For $u_\ell$ one gets
\begin{eqnarray}
\label{def-u}
 u_\ell(r) = F_l(\eta, kr), \ \ r\le R, \nonumber \\
 u_\ell(r) = a_1 \hat{j}_l(kr) + b_1 \hat{n}_l(kr), \ \ r> R.
\end{eqnarray}
For $v_\ell$ the solution takes the form
\begin{eqnarray}
\label{def-v}
 v_\ell(r) = a_2F_\ell(\eta, kr)+b_2G_\ell(\eta, kr), \ \ r\le R\nonumber \nonumber \\
 v_\ell(r) = \hat{h}^+_\ell(kr), \ \ r> R.
\end{eqnarray}
In these representations the Zommerfeld  parameter $\eta$ is defined by the standard expression $\eta =V_0/(2k)$. By  $\hat{j}_l$, $\hat{n}_l$ and $\hat{h}^+_\ell$ we denote
the Riccati-Bessel, Riccati-Neumann and Riccati-Hankel functions which are related
to the respective spherical Bessel functions as for example $\hat{j}_l(z)=z{j}_l(z)$. For spherical Bessel functions and  for the regular $F_\ell$ and irregular $G_\ell$ Coulomb functions we use the normalization of \cite{messiah}.
Coefficients $a_1$, $b_1$, $a_2$ and $b_2$ should be determined in such a way that both functions $u_\ell$, $v_\ell$ and their first derivatives are continuous at $r=R$.
This yields
\begin{eqnarray}
\label{abcoef}
 a_1 =  -W_R(F_\ell,\hat{n}_\ell)/k,\quad b_1 =  W_R(F_\ell,\hat{j}_\ell)/k, \nonumber \\
 a_2 = -W_R(\hat{h}^+_\ell,G_\ell)/k,\quad b_2 = W_R(\hat{h}^+_\ell,F_\ell)/k,
\end{eqnarray}
where $W_R$ stands for Wronskian that is calculated at $r=R$. Now the Wronskian $W(u_\ell,v_\ell)$ from (\ref{Green's uv})
can  easily be computed and takes the form
\begin{equation}
\label{wronsk}
W(u_\ell,v_\ell) = W_R(F_l,\hat{h}^+_\ell).
\end{equation}
Equations~(\ref{Green's uv}) -- (\ref{wronsk}) completely  determine the partial Green function $G_{R\ell}$.

For our needs of evaluating the short-range asymptotic behavior of the Green function $G_R$, if $R$ is well separated from zero,
the region should be considered where $r<R $ and $r'<R$ .
In this case, by inserting into (\ref{Green's uv})  the quantities calculated above  we finally represent the partial Green function by the sum of two terms
\begin{equation}
\label{final}
 G_\ell(r,r',k^2) = \frac{1}{k}F_\ell(\eta, kr_<)H^+_\ell(\eta, kr_>) + \frac{\chi_{R\ell}(k)}{k} F_\ell(\eta, kr) F_\ell(\eta, kr'),
\end{equation}
where $\chi_{R\ell}(k)$ is given by
\begin{equation}
 \chi_{R\ell}(k) =- W_R(\hat{h}^+_\ell,H^+_\ell)/W_R(\hat{h}^+_\ell,F_\ell).
 \label{zeta}
\end{equation}
Here the Coulomb outgoing wave $H^+_\ell$ is introduced according to the definition  $H^+_\ell=G_{\ell}+\mbox{i}F_{\ell}$.

Now we have all components  which are needed for calculating the Green function $G^+_R$ by the formula (\ref{series}).
In view of (\ref{final}) the representation for $G^+_R$ in the region where $r,r' < R$ is given by the sum of two terms
\begin{equation}
\label{3d final}
 G^+_R(\bm{r},\bm{r'},k^2) = G_{\cal C}(\bm{r},\bm{r'},k^2+\mbox{i}0) + Q_R(\bm{r},\bm{r'},k^2).
\end{equation}
The first term is the conventional Coulomb Green function which is calculated by the partial wave series \cite{newton} as
\begin{equation}
G_{\cal C}(\bm{r}, \bm{r'}, k^2+\mbox{i}0)= \frac{1}{4 \pi k}\sum_{\ell=0}^{\infty}(2\ell+1)
\frac{ F_\ell(\eta, kr_<)H^+_\ell(\eta, kr_>)}{rr'}
P_\ell(\bm{\hat{r}} \cdot \bm{\hat{r}'}).
\label{CoulombGFpartial}
\end{equation}
The second term $Q_R$ reads
\begin{equation}
\label{tildeg}
 {Q_R}(\bm{r},\bm{r'},k^2) = \frac{1}{4\pi k} \sum_{\ell=0}^{\infty} (2\ell+1)
 \chi_{R\ell}(k)
 \frac{F_\ell(\eta, kr)F_\ell(\eta, kr')}{rr'}
 P_\ell(\bm{\hat{r}} \cdot \bm{\hat{r}'}).
\end{equation}
The last formula can be rewritten in terms of the Coulomb Green functions taken on the
upper and lower rims of the cut along the positive real semi axis of the energy complex plane. This can be made by using the formula
$F_\ell=(H^{+}_{\ell}-H^{-}_{\ell})/(2\mbox{i})$ and the method of the paper \cite{yak-screened-coulomb}. The result reads
\begin{equation}
Q_R(\bm{r},\bm{r}') = \frac{1}{2\mbox{i}} \int_{-1}^{1}\mbox{d}\zeta \,  Z_R(\xi,\zeta)
[G_{\cal C}(r,r',\zeta,k^2+\mbox{i}0)-G_{\cal C}(r,r',\zeta,k^2-\mbox{i}0)].
\label{QQ}
\end{equation}
Here the parameter $\xi$ is defined as $\xi=\hat{ \bm{r}}\cdot \hat{ \bm{r'}}$. The kernel $Z_R$ is given by the decomposition
in Legendre polynomials
\begin{equation}
Z_{R}(\xi,\zeta) = \sum_{\ell=0}^{\infty} (\ell+1/2)\chi_{R\ell}(k)P_{\ell}(\xi)P_{\ell}(\zeta).
\label{Z_R}
\end{equation}
As it may be seen from the Hostler representation \cite{hostler}, the Coulomb Green function $G_R(\bm{r},\bm{r'},z)$  actually depends on $r$, $r'$ and the angle between vectors $\bm{r}$ and $\bm{r'}$ through the expression $\hat{ \bm{r}}\cdot \hat{ \bm{r'}}$. We have reflected this fact in notations in  the integrand of (\ref{QQ}) where  $\zeta$ stands for $\hat{ \bm{r}}\cdot \hat{ \bm{r'}}$.
The formulae (\ref{final}-\ref{Z_R}) are valid only in that part of configuration space where $r,r' < R$. Nevertheless, this representation of the Green function in that region completely defines the transition operator $T_R(z)=V_R-V_RG_R(z)V_R$ . Indeed, with (\ref{3d final}) the operator $T_R$ takes   the form
\begin{equation}
T_R(z)=V_R-V_RG_{\cal C}(z)V_R - V_RQ_R(z)V_R.
\label{T}
\end{equation}
It is interesting to see what is the consequence of (\ref{T}) when $R\to \infty$. Evaluating the right hand side of (\ref{zeta})  asymptotically  when $kR \gg \ell(\ell+1)$ and $kR \gg \ell(\ell+1)+\eta^2$~\cite{abram} we come to the expression for $\chi_{R\ell}(k)$
\begin{equation}
 \chi_{R\ell}(k) =  \mbox{i}\eta \exp(2\mbox{i}\theta_\ell)/(k R) + {\cal{O}}(1/R^2),
\label{zeta ass}
\end{equation}
where $\theta_\ell = kR - \eta\log(2kR) - \pi\ell/2 + \sigma_\ell$ and $\sigma_\ell = \arg \Gamma(\ell+1+\mbox{i}\eta)$
is the Coulomb phase shift. From (\ref{Z_R}) we have for the $L_2(-1,1)$ norm of the kernel $Z_R$
\begin{equation}
\|Z_R\| = \max\limits_{\ell} |\chi_{R\ell}(k)| = \eta/(kR) + {\cal O}(R^{-2}).
\label{Z_R norm}
\end{equation}
Hence, the last term in (\ref{T}) is negligible when $R\to \infty$ and therefore
\begin{equation}
T_R(z)= V_R-V_RG_{\cal C}(z)V_{R} + {\cal{O}}(R^{-1}).
\label{T_R ass}
\end{equation}
This can be used for calculating the limit of $T_R$ when $R\to \infty$ and it will be done in another publication.

In the last part of this section we calculate the asymptote of $G^+(\bm{r},0, k^2)$ when $r\to 0$. We start from the second term in
(\ref{3d final}). Since \cite{abram}
\begin{equation}
 F_\ell(\eta, x) = C_\ell(\eta) x^{\ell+1} \left( 1 + \eta x/(\ell+1) + {\cal{O}}(x^2) \right),
\end{equation}
as $x \to 0$, the leading order behavior of $Q_R$ reads
\begin{equation}
Q_{R}(\bm{r},0,k^2)= C_0(\eta)\chi_{R0}(k)F_{0}(\eta,kr)/(4\pi r)+ {\cal{O}}(kr)
\label{Q_R zero}
\end{equation}
as $kr \to 0$. From this it is seen that $Q_{R}(\bm{r},0,k^2)$ has the finite limit
\begin{equation}
\lim_{r\to 0}{Q_{R}(\bm{r},0,k^2)}=k C^2_0(\eta)\chi_{R0}(k)/(4\pi),
\label{lim Q}
\end{equation}
where $C_0^2(\eta)=2\pi\eta(e^{2\pi\eta}-1)^{-1}$. This analysis shows us that the singular behavior of $G^+_R(\bm{r},0,k^2)$ at small $r$  comes exclusively from the first term in (\ref{3d final}), i.e. from the Coulomb Green function, which as $r\to 0$ has the asymptote \cite{yak12}
\begin{equation}
\label{gfasym}
G_{\cal C}(\bm{r},0,k^{2}+\mbox{i}0) = \frac{1}{4\pi}\left[
1/r+V_0\log r\right]+C(k)+{\cal{O}}(r\ln r).
\end{equation}
Here $C(k)$ is given by
\begin{equation}
C(k)=\frac{\mbox{i}k}{4\pi}+\frac{V_0}{4\pi}\left[ \log(-2\mbox{i}k)+\psi\left(1+\mbox{i}\eta\right)+2\gamma_{0}-1 \right] ,
\label{C(k)}
\end{equation}
where $\gamma_0$ is the Euler-Mascheroni constant and $\psi(z)$ is the digamma function \cite{abram}.

Collecting together the expressions obtained above we arrive at the following result on the short-range behavior of the Green function $G^+_R$
\begin{equation}
G^+_R(\bm{r},0,k^2) =  \frac{1}{4\pi}
\left[
1/r+V_0\log r\right] +C(k)+ \frac{k C^2_0(\eta)\chi_{R0}(k)}{4\pi}+{\cal{O}}(r\ln r).
\label{G_R ass}
\end{equation}
The latter finalizes our study of Green's function for the sharply screened Coulomb potential. It is apparent that in this case  the zero-range potential will  be identical to that of the case of the pure Coulomb potential \cite{yak12}.

\section{Green's function behavior in the case  of $\mathfrak{V}(\rho,\delta)$ class potentials}
\label{anypot}

In this section we study the Green function short-range asymptote  for potentials of the $\mathfrak{V}(\rho,\delta)$~class.
The Lippmann-Schwinger integral equation
\begin{equation}
G^+(\bm{r}, \bm{r'},k^2)=G^+_{0}(\bm{r},\bm{r'},k^2)- \int \mbox{d}\bm{q}\, G^+_{0}(\bm{r},\bm{q},k^2) V(\bm{q}) G^+(\bm{q}, \bm{r'},k^2)
\label{LS G}
\end{equation}
with
\begin{equation}
G^+_{0}(\bm{r},\bm{r'},k^2)=\frac{1}{4\pi} \frac{\exp(\mbox{i}k|\bm{r}-\bm{r'}|)}{|\bm{r}-\bm{r'}|}
\label{G_0}
\end{equation}
in this case has the unique solution \cite{newton, povzner} which completely defines the Green function $G^+(k^2)$ . For our purpose of evaluating the short-range behavior we set $\bm{r'}=0$ and iterate (\ref{LS G}) one time  which results
\begin{eqnarray}
G^+(\bm{r}, 0,k^2)=G^+_{0}(\bm{r},0,k^2)-
\int \mbox{d}\bm{q}\, G^+_{0}(\bm{r},\bm{q},k^2) V(\bm{q}) G^+_0(\bm{q}, 0,k^2)\nonumber \\
+ \int \mbox{d}\bm{q}\, G^+_{0}(\bm{r},\bm{q},k^2) V(\bm{q})
\int \mbox{d} \bm{q'} G^+_0(\bm{q}, \bm{q'},k^2)V(\bm{q'}) G^+(\bm{q'}, 0,k^2).
\label{LS G0}
\end{eqnarray}
Now we consecutively  consider the short-range behavior of right hand side terms. The first term asymptote as $r\to 0$ is obvious
\begin{equation}
G^+_{0}(\bm{r},0,k^2)=\frac{1}{4\pi r} + \frac{\mbox{i} k}{4\pi} + {\cal O}(r).
\label{G_0 ass}
\end{equation}
For evaluating the second term on the right hand side of (\ref{LS G0}) it is useful to split the integral into two parts in order
to separate the short-range and long-range contributions of the integrand.
Let us consider the integrals
\begin{equation}
I_j(\bm{r})=\int_{\Omega_j} \mbox{d}\bm{q}\, G^+_{0}(\bm{r},\bm{q},k^2) V(\bm{q}) G^+_0(\bm{q}, 0,k^2)
\label{I omegaj}
\end{equation}
over domains $\Omega_j\in \mathbb{R}^3$ defined as $\Omega_{1(2)}=\{\bm{q}: q<(>)r_0\}$. The radius $r_0$
can be chosen as any positive bounded number well separated from zero and will be specified below.

For the integral $I_1(\bm{r})$ we can expand the Green function factors of the integrand by using the Taylor decomposition up to quadratic terms  as
\begin{equation}
G^+_0(\bm{r},\bm{q},k^2)= 1/(4\pi|\bm{r}-\bm{q}|) + \mbox{i}k/(4\pi) + {\cal O}(|\bm{r}-\bm{q}|^2)
\label{G_0 exp}
\end{equation}
and similar expression holds for $G^+_0(\bm{q},0,k^2)$ with $\bm{r}$ set to $0$. For the potential $V(\bm{q})$ in $\Omega_1$ we assume that $r_0$ is chosen such that the formula (\ref{V at or}) can be used and
the $W(\bm{q})$ factor can be given by its Taylor decomposition
\begin{equation}
W(\bm{q})=V_0 + \bm{q}\cdot \nabla W(0) + {\cal O}(q^2).
\label{W}
\end{equation}
Then the most singular term of $I_1(\bm{r})$ as $r\to 0$ will be generated by introducing in (\ref{I omegaj}) with $j=1$
the leading terms of integrand constituents which are  defined by (\ref{G_0 exp}) and (\ref{W}). This will lead us to the integral
\begin{equation}
I^s_1(\bm{r})= V_0/(4\pi)^2 \int_{\Omega_1} \mbox{d} \bm{q}\,|\bm{r}-\bm{q}|^{-1}\, q^{-\rho-1}.
\label{Is}
\end{equation}
Since we need to find the short-range behavior of this integral when $r\to 0$ we can always take $r$ such that $r< r_0$. In this case the evaluation of the integral in (\ref{Is}) is easy to perform  with the help of the formula
\begin{equation}
 \frac1{|\bm{q} - \bm{q'}|} = \frac{1}{q_>} \sum_{\ell=0}^\infty  \frac{q_<^\ell}{q_>^{\ell}}\,
 P_\ell\left(\bm{\hat{q}}\cdot\bm{\hat{q}'}\right),
\label{gen}
\end{equation}
where as usual $q_>=\max\{q,q'\}$ and $q_<=\min\{q,q'\}$. The results are naturally separated in two ones, i.e. if $\rho\ne 1$ it reads
\begin{equation}
I^s_1(\bm{r})= \frac{V_0}{4\pi(2-\rho)(\rho -1)} r^{-\rho + 1} + \frac{V_0 }{4\pi (1-\rho)}r_0^{-\rho+1}
\label{Is res}
\end{equation}
and when $\rho=1$ the integral $I^s_1$ takes the form
\begin{equation}
I^s_1(\bm{r})= - \frac{V_0}{4\pi}\log(r) + \frac{V_0}{4\pi}[1+\log(r_0)].
\label{Is res1}
\end{equation}
From (\ref{Is res}) it is seen that if $1<\rho<2$ then $I^s_1$ has the polar singularity $r^{-\rho+1}$ whereas if $\rho<1$ the first term in (\ref{Is res}) is vanishing as $r\to 0$ and $I^s_1$ is regular and has a finite limit. From this analysis  of  the integral (\ref{Is}) it becomes
clear that taking into account the less singular terms in the expressions for the integrand of the integral $I_1(\bm{r})$  one will obtain  non singular contributions as $r\to 0$.

Let us now consider the integral $I_2(\bm{r})$. For the  modulus of $I_2(\bm{r})$ we can easily arrive at the inequality
\begin{equation}
|I_2(\bm{r})| \le \frac{1}{(4\pi)^2}\int_{\Omega_2} \mbox{d} \bm{q}\, \frac{|V(\bm{q})|}{q\, |\bm{r}-\bm{q}|}.
\label{I2}
\end{equation}
Let us now suppose that $r_0$ is chosen such that the inequality (\ref{Vbound}) can be used for $q>r_0$
then with the help of (\ref{gen}) the right hand side of (\ref{I2}) can be estimated as
\begin{equation}
 \int_{\Omega_2} \mbox{d} \bm{q}\, \frac{|V(\bm{r})|}{q\, |\bm{r}-\bm{q}|}
 \le
 C \int_{r_0}^{\infty} \mbox{d}q\, (1+q)^{-1-\delta}.
 \label{integ}
 \end{equation}
 Since the last integral converges, the integral $I_2(\bm{r})$ is uniformly bounded  for all $\bm{r} $  such that $r<r_0$.

It remains to estimate the last third term in (\ref{LS G0}). The inner integral over $\bm{q'}$ by its structure is quite similar to the integrals considered above if $G^+_0(\bm{q'},0,k^2)$ stands instead of $G^+(\bm{q'},0,k^2)$.  In this case the inner integral as the function of $\bm{q}$ may have a singularity that is not stronger than $q^{-\rho+1}$ one. As it has been already shown  such a  singularity in the integrand of the outer integral over $\bm{q}$ will lead to the nonsingular behavior of the result as the function of $\bm{r}$ in the vicinity of the point $\bm{r}=0$. On using the iterative arguments 
this result can easily be extended on the case of the genuine integrand in the third term of (\ref{LS G0}) \cite{povzner}.  Thus, the last term in (\ref{LS G0}) should  have a finite limit as $r\to 0$.

Collecting the results obtained in this section we formulate the final  statement about the short-range behavior of the function $G^+(\bm{r},0,k^2)$ for the case of $1<\rho<2$ as
\begin{equation}
G^+(\bm{r},0,k^2) = \frac{1}{4\pi} \left[1/r  +A_0/r^{\rho-1} \right] + B_1 +{ o}(1),
\label{rez 1}
\end{equation}
for the case of $\rho=1$
as
\begin{equation}
G^+(\bm{r},0,k^2) = \frac{1}{4\pi} \left[1/r  +V_0\log(r) \right] + B_{2} +{ o}(1),
\label{rez 2}
\end{equation}
and for the case of $\rho<1$ as
\begin{equation}
G^+(\bm{r},0,k^2) = \frac{1}{4\pi r}  + B_{3} +{ o}(1).
\label{rez 3}
\end{equation}
Here the constant $A_0$ is given by
$$
A_0=\frac{V_0}{(2-\rho)(1-\rho)}
$$
and  all finite contributions from respective integrals are denoted by $B_j$, $j=1,2,3$.

\section{Application to zero-range potential formalism}
The zero-range potential is introduced by implementation of a special singular boundary condition on the
solution of the Schr\"odinger equation at small inter-particle distances \cite{albev}. This boundary condition
can be enforced  \cite{yak12} if the delta functional term is inserted into equation (\ref{SEVdelta}). In this case the solution is represented
by
\begin{equation}
\phi(\bm{r},\bm{k}) = \phi_0(\bm{r},\bm{k}) - \lambda G^+(\bm{r},0,k^2)\beta.
\label{psi-LSS}
\end{equation}
Here $\phi_0$ is defined according to (\ref{SEV}, \ref{asym bc}). Since the asymptotic behavior of Green's function has been studied in details in preceding section it remains to estimate the short range behavior of $\phi_0$. It
can be done on the basis of the subsequent Lippmann-Schwinger integral equation
\begin{equation}
\phi_0(\bm{r},\bm{k})=\exp(\mbox{i} \bm{k}\cdot \bm{r}) - \int \mbox{d}\bm{q}\, G^+_{0}(\bm{r},\bm{q},k^2) V(\bm{q}) \phi_0(\bm{q},\bm{k}).
\label{LS psi}
\end{equation}
It is easy to see that for a potential $V\in \mathfrak{V}(\rho,\delta)$  with $\rho<2$, $\delta>1$ any iteration of (\ref{LS psi}) is bounded
function with a finite limit at $\bm{r}=0$ and so is the solution. For completeness let us prove  this statement for the first iteration which we represent as the sum of two integrals $J_1$ and $J_2$ defined by
\begin{equation}
J_j(\bm{r})= \int_{\Omega_j} \mbox{d}\bm{q}\, G^+_{0}(\bm{r},\bm{q},k^2) V(\bm{q}) \exp(\mbox{i}\bm{q}\cdot \bm{k}), \ \ j=1,2.
\label{1 iter}
\end{equation}
As in the previous section, we can obtain  the following estimate for the modulus of $J_2(\bm{r})$ for all $r$ such that $r<r_0$
\begin{equation}
|J_2(\bm{r})| \le
\frac{C}{4\pi}\int_{\Omega_2}\mbox{d}\bm{q}\   q^{-1}(1+q)^{-1-\delta}.
\label{J 2}
\end{equation}
The integral on the right hand side of (\ref{J 2}) converges. Thus,  $J_2(\bm{r})$ is uniformly bounded in the neighborhood of the origin.
The integral $J_1(\bm{r})$ can be treated  by similar method which was used in the previous section for the integral $I_1(\bm{r})$. Particularly, if the most singular terms in the integrand are kept then the leading order of the integral $J_1(\bm{r})$ takes the form
\begin{equation}
J_1(\bm{r})\sim \frac{V_0}{4\pi} \int_{\Omega_1} \mbox{d} \bm{q}\, |\bm{r}-\bm{q}|^{-1}  q^{-\rho}.
\label{J 2 ass}
\end{equation}
The integral on the right hand side of this formula coincides  with the integral from (\ref{Is}) if  $\rho+1$ in (\ref{Is}) is replaced by $\rho$. Then from the analysis made in the previous section it follows that the integral $J_1(\bm{r})$ is regular as $r\to 0$.
As the result  we can conclude  that for the potentials of the class $\mathfrak{V}(\rho,\delta)$  the
short-range asymptote of the wave function $\phi_0(\bm{r}, \bm{k})$ is regular as $r\to 0$  and  $\phi_0(\bm{r}, \bm{k})$ has a finite limit at the origin.

The singular behavior of the full solution $\phi$ is therefore completely determined by the singularities of the Green function. Accordingly, we have three following cases of wave function asymptotes depending on the value of $\rho$. If $2>\rho>1$ the asymptote is of the form
\begin{equation}
\phi(\bm{r},\bm{k}) = \frac{\alpha_1}{4\pi} \left[1/r  +A_0/r^{\rho-1} \right] + \beta_1 + { o}(1).
\label{rez 11}
\end{equation}
For the case of $\rho=1$ it is given by
\begin{equation}
\phi(\bm{r},\bm{k}) = \frac{\alpha_2}{4\pi} \left[1/r  +V_0\log(r) \right] + \beta_2 +
{ o}(1),
\label{rez 22}
\end{equation}
and for the case of $\rho<1$ the asymptote reads
\begin{equation}
\phi(\bm{r},\bm{k}) = \frac{\alpha_3}{4\pi r}  + \beta_3 +{ o}(1).
\label{rez 33}
\end{equation}
Here the constants are given by following expressions
$$
A_0=\frac{V_0}{(2-\rho)(1-\rho)},
$$
$\alpha_j=-\lambda \beta_j$ and $\beta_j=\phi_0(0,\bm{k})-\lambda B_j\beta_j$ for  $j=1,2,3$. The case of $\rho=1$ completely coincides
with the case of the zero-range potential for the  Coulomb potential \cite{yak12}.

Alternatively, the zero range potential  is defined by the pseudo-potential. Following the procedure described in the paper \cite{yak12}
the form of the pseudo-potential is calculated from the asymptotic expansions (\ref{rez 11}-\ref{rez 33}). For all three cases
the pseudo-potential may be given by the uniform representation
\begin{equation}
\lambda W_j(\bm{r}) = \lambda \delta(\bm{r}) \frac{\mbox{d}}{\mbox{d}\omega_j } \omega_j
\label{psp}
\end{equation}
with variables $\omega_j$ determined by formulae
\begin{eqnarray}
1/\omega_1= r^{-1} + A_0r^{-\rho+1},  \\
1/\omega_2= r^{-1} +V_0\log(r), \\
1/\omega_3=r^{-1}.
\end{eqnarray}

\section{Conclusion}
We have shown that the Green function of a Hamiltonian with the potential possessing singular behavior ${\cal O}(r^{-\rho})$ at small
inter-particle distances receives the additional singularity ${\cal O}(r^{-\rho+1})$ except for the case of $\rho=1$ when
the logarithmic singularity appears. We have considered the class $\mathfrak{V}(\rho,\delta)$ of potentials with $\rho<2$ and $\delta>1$.
This choice of parameters is not exhausting. With little effort the weaker condition $\delta>0$ can easily be implemented in the theory by performing more delicate estimations for integrals $I_2$ and $J_2$ which should be treated as non absolutely convergent integrals.
We have left out such an analysis in order not to make the paper too long.
\ack
This work was partially supported by St Petersburg State University under project No. 11.0.78.2010.
The support by the Ministry of Education and Science of the Russian Federation under project No. 2012-1.5-12-000-1003-016 is also acknowledged.

\section*{References}

\bibliographystyle{unsrt}

\end{document}